# Quantifying Changes to Healthcare Utilization After a Reduction in Cost-sharing Among Deductible Plan Enrollees

Kris Wain, PhD[1]

Debra P. Ritzwoller, PhD[1]

Marcelo Coca Perraillon, PhD[2]

[1]Institute for Health Research, Kaiser Permanente Colorado, Aurora, CO

[2]Department of Health Systems, Management and Policy, University of Colorado Anschutz Medical Campus, Aurora, CO

**Corresponding Author:**

Kris Wain, PhD
Institute for Health Research
16601 East Centretech Parkway
Aurora, CO 80011
Office: 303-636-2435
e-mail: kris.f.wain@kp.org

**ABSTRACT**

Health plan deductibles are a form of cost-sharing that require patients to pay out-of-pocket before insurance pays for benefits. Deductible plans have become increasingly common in the United Sates to mitigate escalating healthcare costs. Quantifying the impact of increased cost-sharing from deductibles on utilization is a challenging empirical question because individuals and employers self-select into plans with deductibles. We evaluated the impact of cost-sharing in plans with deductibles by leveraging an accidental injury to a family member as an instrumental variable that strongly predicted the non-injured family member reaching their deductible maximum, which resulted in a reduction in cost-sharing. Our outcome measures examined utilization subject to cost-sharing as compared to utilization exempt from cost-sharing. Using data from the same healthcare system to control for quality and provider network, we found that reaching the deductible increased emergency department (ED) utilization by 10.0 percentage points (pp). Nearly one-quarter of the increased ED utilization was potentially avoidable. Wellness visits not subject to cost-sharing decreased by 5.7 pp. Results were similar for high-deductible plans and for families meeting their maximum out-of-pocket amount. These findings provide causal evidence that individuals enrolled in plans with deductibles change utilization patterns after an exogenous reduction in cost-sharing.

**Introduction**

The proliferation of health insurance plans that require cost-sharing in the form of annual deductibles has substantially increased the number of individuals liable for a larger share of their healthcare costs. A deductible is the amount a person or family pays out-of-pocket for covered healthcare services before their insurance plan begins to pay. As of 2023, 90% of commercially insured individuals in the United States faced a general annual deductible ("KFF Employer Health Benefits Survey Section 7: Employee Cost Sharing" 2023). Some plans are categorized as high-deductible health plans (HDHPs), defined by 2023 IRS guidelines as an individual deductible greater than or equal to $1,500, or a family deductible greater than or equal to $3,000 (IRS 2023). In 2023 nearly one-third of commercially insured individuals were enrolled in a HDHP, while over two-thirds of large employer groups offered a HDHP option ("KFF Employer Health Benefits Survey Section 8: High-Deductible Health Plans with Savings Option" 2023).

Health plans with deductibles are designed to incentivize patients to efficiently consume healthcare by sharing costs between the individual and insurer (Robinson 2002). Once the annual deductible amount is met, the cost for future healthcare is reduced to a co-insurance percentage of the total cost or a copay amount. As described in Figure 1, individuals will continue to pay the co-insurance or copay until they reach a maximum out-of-pocket amount, after which covered services become exempt from cost-sharing. Proponents argue that higher cost-sharing leads to efficient healthcare consumption because individuals are aware of treatment costs, limiting unnecessary utilization (Herzlinger 2002; Kolasa and Kowalczyk 2016). On the other hand, those opposed argue that increased cost-sharing leads to reductions for all types of healthcare, resulting in missed opportunities for high-value and preventive care (Brot-Goldberg, Chandra, Handel, and Kolstad 2017). Furthermore, patients may misunderstand their plan benefits, forgoing preventive care that is not subject to cost-sharing (Loewenstein et al. 2013).

In recent years, individuals have been "nudged" into deductible plans with higher cost-sharing to stem the ballooning cost of healthcare, which accounted for an estimated 18% of the United States GDP spending in 2023 (Martin, Hartman, Washington, and Catlin 2019; Maxwell and Temin 2002). To incentivize enrollment into employer-sponsored deductible plans, individuals are typically offered lower (or zero dollar) monthly premiums and may be provided with health reimbursement arrangements (HRA) or Flexible Spending accounts (FSA), employer-sponsored spending accounts providing first-dollar coverage for qualifying healthcare services. Individuals enrolled in IRS qualified HDHPs are also eligible for a health savings account (HSA), a federally tax-exempt account that can be funded by the employer or employee. HSA accounts can be used to cover specified medical expenses, and the funds remaining at the end of the benefit period can be carried over into subsequent years while accumulating interest.

To encourage utilization of high-value care, the Affordable Care Act (ACA) extended coverage to many forms of preventive care with no cost-sharing requirements regardless of health plan type. The prevention provision of the ACA includes several types of cancer screening, wellness visits, and other forms recommended preventive care that received an A or B recommendation by the US Preventive Services Task Force (Sommers and Wilson 2012). Appropriate use of preventive care increases the likelihood of earlier diagnosis for many chronic health conditions, improves health outcomes, and decreases long-term treatment costs (Seely and Alhassan 2018; Siu and Force 2016; Sabik and Adunlin 2017; Schiffman, Fisher, and Gibbs 2015; Jiang, Hughes, and Wang 2018; Ehreth 2003). However, individuals enrolled in deductible plans may not understand the ACA allows for preventive care with no cost-sharing. In addition, individuals may worry about out-of-pocket costs from follow-up care or high treatment price tags following a diagnosis of an acute or chronic disease (Loewenstein et al. 2013; Wharam et al. 2018). The combination of misunderstood benefits and concerns over high diagnostic and treatment costs may cause individuals enrolled in deductible plans to decrease utilization of recommended preventive care even though they have no out-of-pocket cost requirement.

Quantifying the impact increased cost-sharing from deductibles has on healthcare utilization is a difficult empirical problem because individuals and employers may select into plans with deductibles. Individuals may choose a deductible plan to reduce their monthly premium costs, or because they prefer to utilize less healthcare, even forgoing high-value preventive care that is not subject to a deductible. Moreover, deductible plans may offer wider networks than no-deductible plans, which may incentivize individuals who prefer a broader choice of providers and specialists to select deductibles plans (Berki and Marie 1980). On the other hand, individuals who select plans with no deductibles may prefer lower cost-sharing options because they expect to use larger amounts of healthcare due to existing chronic or other conditions. Thus, cost-sharing, preferences, and health status affect both plan selection and utilization (Cutler and Zeckhauser 1998). In recent years, more employers are exclusively offering plans with deductible, leaving employees with no choice for their deductible levels (*2020 Employer Health Benefits Survey* 2020). Most observational datasets do not capture the plan choice set faced by individuals covered under employer sponsored insurance.

Previous studies have explored the impact cost-sharing has on healthcare utilization. The RAND Health Insurance Experiment found that individuals enrolled in high cost-sharing plans reduced consumption for all types of healthcare services (Aron-Dine, Einav, and Finkelstein 2013). More recently, studies evaluating the impact of increased cost-sharing have reported conflicting findings, particularly regarding preventive care utilization, which is typically exempt from cost-sharing (Khushalani et al. 2020; Rowe, Brown-Stevenson, Downey, and Newhouse 2008; Wharam et al. 2011; Wharam et al. 2012; Johnson et al. 2015; Agarwal, Mazurenko, and Menachemi 2017; Mazurenko, Buntin, and Menachemi 2019; Borah, Burns, and Shah 2011; Brot-Goldberg, Chandra, Handel, and Kolstad 2017). Contemporary studies often compare utilization before and after the implementation of the ACA, a time period during which several types of preventive care transitioned from requiring cost-sharing to being offered with no cost-sharing obligations for eligible individuals, irrespective of their health plan (Sabik and Adunlin 2017; Cooper et al. 2017; Adams et al. 2018). Other evidence relies upon employers who mandated their employees switch from a low cost-sharing plan to a plan with high cost-sharing requirements (Brot-Goldberg, Chandra, Handel, and Kolstad 2017). Studies that rely upon employer mandated health plan switches are unable to explore effects among individuals given a choice of health plans. If employees are aware of the upcoming change to their benefits, they may alter their healthcare utilization patterns based upon anticipated out-of-pocket expenses (Eisenberg et al. 2017). Furthermore, individuals in these studies are typically given the option to select from a wide network of providers. However, most studies are unable to control for the heterogeneous quality of care offered within these large networks (Institute of Medicine Committee on Quality of Health Care in 2000).

The goal of this study is to estimate the causal effect of a reduction in patient cost-sharing on healthcare utilization among individuals enrolled in plans with deductibles. We leverage data from a single healthcare system in which the size of the network and the quality of care is the same for all patients regardless of benefit plan type. To estimate causal effects, we employed an instrumental variable design in which an unexpected accidental injury to a family member increased the likelihood the non-injured family member met their annual family-level deductible, thus decreasing cost-sharing for future healthcare services for the non-injured family member (Kowalski 2016). We focused on two different types of utilization, preventive care that is typically exempt from cost-sharing, and other healthcare services that are typically subject to cost-sharing. During the study period, several types of cancer screenings, wellness visits, and other types of preventive care did not have any cost-sharing requirements for individuals meeting eligibility criteria. Other types of healthcare services such as emergency department (ED) visits, primary care encounters, inpatient stays and imaging were typically subject to cost-sharing. Thus, we ask a different question than prior studies examining the ACA or employer-mandated switches to deductible plans. Rather than assessing the average treatment effect for a cohort following an anticipated change in cost-sharing, we examine how an unexpected and exogenous decrease in cost-sharing (after reaching deductible maximum) affects subsequent healthcare utilization.

Our findings show that a reduction in cost-sharing increased emergency department utilization and reduced utilization of wellness visits. These findings suggest individuals modify their healthcare utilization patterns based upon their out-of-pocket costs, increasing certain utilization that is subject to a cost-sharing, while decreasing certain types of preventive care that is exempt from a cost-sharing. Our research contributes to the literature exploring the effect of cost-sharing on healthcare utilization among individuals enrolled in plans with deductibles.

**Institutional Details for Kaiser Permanente Colorado**

Kaiser Permanente Colorado (KPCO) is a capitated healthcare model in which a premium is paid to access a network of healthcare providers (McHugh, Aiken, Eckenhoff, and Burns 2016). KPCO operates as an integrated healthcare system that serves as both insurer and the principal care delivery system. Once enrolled, an individual "bonds" with a Primary Care Provider (PCP) who serves as the primary point of care coordination for healthcare services. Research indicates that healthcare provided by integrated healthcare systems is linked to increased quality of care as compared to traditional practice management found in open networks (Reiss-Brennan et al. 2016). Importantly, during our study period all individuals enrolled within KPCO had access to the same network of providers regardless of their health plan type, implying provider quality and care networks are held constant across all members.

Membership within the KPCO population is dynamic; individuals may enroll or dis-enroll for numerous reasons such as changes to their employer-sponsored health insurance, open enrollment periods, or certain life events such as marriage or childbirth. During our study period, the profile of the KPCO enrolled population remained stable. The total enrolled population was approximately 650,000 individuals. Annual deductible amounts varied by benefit plan, but the family-level deductible was usually set to twice the amount of the individual deductible. The KPCO population consisted of 47% enrollees on health plans with no annual deductible, including 21% of the population enrolled Medicare Part C plans. Deductible plans comprised 53% of the population, including 22% of the population enrolled in HDHPs. The average family deductible was $3,784, while the average co-insurance was 21 percent. Copays amounts varied by benefit type, with an average of $22 for outpatient visits, $76 for specialty care, $199 for emergency department visits, and $405 for inpatient admissions.

**Conceptual Framework**

The standard model of demand for insurance assumes that individuals who have health plan options choose their plan based upon a trade-off between lower insurance premiums and increased out-of-pocket costs from cost-sharing. When offered a choice of plans, individuals who expect to consume larger amounts of healthcare select plans with higher premiums and lower cost-sharing, while individuals who expect to consume less healthcare select plans with lower premiums and higher cost-sharing (Wilson et al. 2009; Beeuwkes Buntin, Haviland, McDevitt, and Sood 2011; Meyers, Rahman, and Trivedi 2022; Haeder, Weimer, and Mukamel 2015).

Our primary research question asks whether a decrease in patient cost-sharing after meeting the family deductible changes utilization for preventive healthcare exempt from cost-sharing as compared to healthcare services subject to cost-sharing. For recommended preventive care that is exempt from cost-sharing, a decreased cost-sharing structure should have no impact on utilization. However, individuals may misunderstand their benefits, forgoing high-value care exempt from cost-sharing or have concerns about costs resulting from downstream diagnostic work-ups and treatment (Reed et al. 2012; Loewenstein et al. 2013). On the other hand, a decreased cost-sharing structure for healthcare services subject to cost-sharing could have a larger impact on utilization. For example, a patient might postpone addressing their healthcare needs when faced with the obligation to cover 100 percent of their healthcare expenses while under their deductible. However, the same patient may choose to schedule a visit once the family deductible has been fulfilled and out-of-pocket costs are substantially reduced. Alternatively, preferences associated with selecting a deductible plan may dominate the effect of decreased cost-sharing. In this scenario, we would expect no change in consumption for any type of healthcare service after the family deductible has been reached. Other mechanisms are possible. For example, a family's budget for medical expenditures may be exhausted following an expensive injury, restricting expenditures on future healthcare (Collins, Gunja, Doty, and Buetel 2015). Additionally, the non-injured family member may serve as a caregiver to the injured family member and become more engaged with the health system, providing additional opportunities to address their own healthcare needs (Son et al. 2007). Alternatively, caregivers may face time constraints due to the additional responsibility of helping the injured family member, limiting their ability to consume healthcare for their own needs (Adelman et al. 2014). In the methods section, we describe how we test for these hypotheses.

**Methods**

*Data*

Data for this study was obtained from the KPCO Virtual Data Warehouse (VDW), a common data model derived from administrative electronic health records (EHR). The VDW contains information on health plan benefits including information on annual deductible amounts, procedural and diagnostic codes, demographic characteristics, and socio-economic status at the census-tract level (Ross et al. 2014). The VDW was supplemented with patient out-of-pocket expenditure information from an EHR accounting system.

**Outcomes**

We examined two outcomes exempt from cost-sharing that were common types of healthcare utilization within the KPCO health system. First, we examined wellness visits, a benefit covered under the ACA once every 12 months. Wellness visits were identified using standardized procedural codes as described in Appendix Table A1. Second, we examined mammography utilization. Women aged 50 to 74 years are eligible for mammography biennially with no cost-sharing obligations per United State Preventive Services Task Force recommendations (Siu 2016). Receipt of mammography was established using standardized procedural codes from EHR radiology reporting. While other forms of cancer screening are exempt from cost-sharing under the ACA, they are less common (e.g. lung cancer screening) or have complex eligibility criteria and screening intervals (e.g. colorectal cancer screening), thus were not considered in this study.

We also explored outcomes that were typically subject to cost-sharing. First, we explored utilization by healthcare setting. Binary flags were created to indicate utilization in each of the following settings: emergency department, inpatient care, urgent care, ambulatory surgery, and outpatient clinic visits. Settings for each encounter were identified using established VDW grouping algorithms (Ross et al. 2014). We next employed an empirically driven approach by examining the 3 most common procedures in our cohort that typically required cost-sharing. These procedures were identified using procedural codes: office visits, venipuncture, and physical therapy visits. The complete list of procedural codes used to identify these outcomes is available in Appendix Table A1.

**Control Variables**

To identify family units, individuals were linked to the primary account subscriber using membership-level data. We allowed family size to vary between each benefit period, accounting for members being added or removed from the insurance plan. Patient expenditures were compiled from an EHR and claims-based system that provided out-of-pocket expenditures and date of service for each claim. Family out-of-pocket expenses were summed over the benefit period. If the accumulated family expenditures met or exceeded the family deductible amount, we assigned an index date for when the family unit moved to a reduced cost-sharing structure for future healthcare consumption.

Accidental injuries were identified using the International Classification of Disease Version 10 (ICD-10) codes for injury, poisoning, and other external consequences as described in Appendix Table A2. We also classified accidental injuries by severity using the Abbreviated Injury Scale (AIS), an algorithm that assigns severity based on a 6-point scale (Clark, Black, Skavdahl, and Hallagan 2018). An AIS score of 1 represents a minor injury, while an AIS score of 6 represents an untreatable injury. To ensure the expenditures associated with the accidental injury had a meaningful impact on the family-level deductible, we limited to injuries treated in an inpatient or emergency department setting, which usually incur significant out-of-pocket costs.

Comorbidity prevalence was defined during the 12 months prior to the index date using the Quan Coding Algorithm (Quan et al. 2005). The total number of comorbidities was summed into a numerical index, individuals with 3 or more comorbidities were grouped together into a single category. We used the Yost index as our measure of socioeconomic status, a weighted linear combination of education, median income, housing value and employment derived from American Community Survey data (Yu, Tatalovich, Gibson, and Cronin 2014). We grouped the Yost index into quintiles, 1 representing most affluent, and 5 representing most deprived. Race and ethnicity categories included Non-Hispanic White, Non-Hispanic Black, Asian, Other Races and Unknown (a category for individuals that did not report their race or ethnicity). Small sample sizes required us to aggregate Native Hawaiian/Pacific Islanders, American Indian/Alaskan Natives, and multi-race populations into a category for Other Races.

Our cohort included adults aged 18 to 64 between January 1, 2016, and December 31, 2019. Eligibility for preventive care varied based on type of care as described above. If a preventive care outcome was utilized prior to the index date, we excluded the observation from our model estimates because the individual would not be eligible to receive the same type of preventive for the remainder of the benefit period. Individuals aged 65 and older were excluded because Medicare Part C plans offered at KPCO were individual-level and have no annual deductible. We required individuals to be enrolled during the entire 12-month benefit period to ensure each family unit had equivalent time to make expenditures that count toward their deductible. Each individuals' expenditures were accumulated toward the family deductible until their individual deductible was met. If the individual deductible was met, the individual's expenditures no longer contributed to the family deductible. To allow the family deductible to be met without requiring the individual deductible to be met, we limited our cohort to families with 3 or more individuals insured on the same plan. Families experiencing 2 or more accidental injuries in the same benefit period were excluded because they may face different barriers to healthcare utilization.

**Empirical Strategy**

A naïve identification strategy would examine utilization outcomes among those who met their family deductible and moved to a reduced cost-sharing structure as compared to those who did not meet their family deductible. However, this naïve setup would likely produce endogenous estimates. To circumvent the endogeneity of plan selection and preferences for healthcare utilization, we exploit the occurrence of an accidental injury to act as a pseudo-randomizer for reaching the family-level deductible during the benefit period. Specifically, an accidental injury occurring to a family member increases the likelihood that the non-injured family member will meet their annual family-level deductible. When the non-injured family member meets their family-level deductible, out-of-pocket costs decrease for future healthcare that is subject to cost-sharing as described in Figure 2.

To estimate the casual effect of a decrease in cost-sharing on healthcare utilization we employ a two-stage residual inclusion (2SRI) methodology, a form of instrumental variable (IV) modeling that allows for consistent estimation of non-linear models. This method is appropriate for our study because our outcomes are binary indicators for healthcare utilization (Terza, Basu, and Rathouz 2008). In contrast to two-stage least squares (2SLS) techniques traditionally used in IV models, 2SRI includes the endogenous regressor in the second stage and includes the first-stage residuals as an additional regressor. In a parametric framework, the 2SRI methodology has been shown to provide more consistent estimates than traditional 2SLS techniques for linear models. As a sensitivity analysis, we estimated 2SLS models which yielded similar results to our 2SRI models (results not shown).

In the first stage, we estimated a logit model for the likelihood of meeting a family deductible, the endogenous treatment variable. We assumed the accidental injury to a family member was exogenous, conditional on covariates, and not related to future healthcare utilization for the uninjured family member. Thus, our design allowed us to identify the causal effect of decreased cost-sharing after a family deductible is met. The first-stage logit regression model was estimated as:

$$\Pr(deductible_{it=1}|injury, \mathbf{z}) = f(\alpha_1 + \alpha_2 injury_{it} + \boldsymbol{\alpha}\mathbf{z}_{it}), \quad (1)$$

Where *i* indexes an individual and *t* indexes the annual benefit period. A distinct observation was recorded for every benefit period in which an individual met inclusion criteria. *Deductible* is a binary flag indicating the family deductible had been reached for individual *i* in benefit period *t*. Because the family deductible resets at the beginning of each benefit period, we allowed *deductible* to dynamically change for each benefit period. *Injury* is a binary indicator for an individual having a family member that experienced an accidental injury in benefit period *t*. Our vector of explanatory variables **z** included the family deductible amount, family size, age of eligible individual, race, ethnicity, comorbid status, and the Yost quintile. From regression (1) we also obtained residuals in the probability scale for each predicted value of meeting the family deductible, $\hat{r}_{it}^{deductible}$.

In the second stage we estimated a logit model for each utilization outcome as a function of meeting a family deductible, other explanatory variables, and residuals from our first stage model. Standard errors were bootstrapped using 1,000 replications and clustered at the family level. The second stage equation was estimated as:

$$\Pr(utilization_{it=1}|deductible, \mathbf{z}, \hat{r})y_{it} = f(\lambda_1 + \lambda_2 deductible_{it} + \lambda_3 \hat{r}_{it}^{deductible} + \boldsymbol{\lambda} \mathbf{z}_{it}), (2)$$

The coefficient of interest in the second stage is $\lambda_2$, the local average treatment effect (LATE) of meeting a family deductible on future utilization.

The main assumption in instrumental variable designs is the exclusion restriction; any effect the instrumental variable had on the outcome only occurred through its effect on the treatment variable. In the context of our study, the exclusion restriction required that the only way a family member's accidental injury influenced the non-injured family member's healthcare utilization was through decreased cost-sharing after meeting the family deductible, conditional on other covariates, and no other unobserved factors associated with the injury and future utilization. We constructed the instrumental variable at the family level, but the outcome is measured at the individual level. Creating the instrument in this way removed the effect of the injured family member's follow-up care from the non-injured family member's outcomes. Additionally, we did not consider utilization for the individual who experienced the accidental injury.

Families with lower deductible amounts might be more likely to meet their annual deductible following an accidental injury and could be more likely to consume healthcare regardless of meeting their deductible for other reasons. To explore this potential heterogeneity, we stratified our analyses by deductible amount (HDHP vs low deductible). It is also possible that certain families engaged in riskier behaviors and were more likely to suffer injuries that required healthcare services. We examined this potential bias by limiting our sample to the first accidental injury that occurred within a family unit. We also performed a sensitivity analysis that excluded individuals who met their family out-of-pocket maximum, which reduced their future healthcare costs to $0 and could lead to different utilization patterns as compared to only meeting a deductible.

Individuals in our cohort who met their family deductible had a natural index date in which to measure healthcare utilization outcomes, beginning on the date their deductible was met and ending the last day of the benefit period. However, the comparison group who did not meet their family deductible did not have a natural index date to measure utilization outcomes. Thus, we generated a random index date for the group of individuals not meeting their family deductible that followed the same distribution of dates observed for individuals who did meet their deductible (Jacob et al. 2020). This random index date provided both groups with equivalent time intervals to measure utilization outcomes.

Another potential threat to the exclusion restriction was a family member's injury indirectly affecting utilization for the non-injured family member other than through a reduction in cost-sharing. We addressed this concern in several ways. The diagnostic codes we examined for accidental injuries were unrelated to family genetic conditions. Injuries linked to genetic conditions within a family might be linked to higher utilization among non-injured family members diagnosed with the same condition, potentially violating the exclusion restriction. The diagnostic categories we included were open wounds, fractures, dislocations, crushing injuries, burns, toxic effects of non-medical substances, and complications of surgical and medical care. While most injuries are promptly resolved with appropriate medical care, severe injuries such as brain trauma, amputation, and spinal cord damage may require additional support from the non-injured family members. These family member caregivers may be more engaged with the healthcare system as they assist with their injured family member's follow-up care needs, providing additional opportunities for the caregiver to address their own care needs. Alternatively, caregivers may spend additional time supporting their injured family member, allowing less time for their own healthcare needs. We explored this potential bias by excluding severe injuries in a sensitivity analysis.

To better understand the mechanisms causing changes in healthcare utilization patterns, we examined heterogeneous effects by age and sex. We estimated 2SRI models stratified by age categories; 18 to 30 years of age when individuals often have less purchasing power, 31 to 50 when individuals may have more financial resources but might also support a family, and 51 to 64 when individuals may have greater purchasing power but might also require care for comorbid health conditions. To examine how reduced cost sharing interacts with gender, we estimated models stratified by sex (male vs female). Additionally, we explored whether ED visits were being used more frequently for non-emergent care after meeting the family deductible by leveraging New York University's Emergency Department Algorithm (NYU-EDA). The NYU-EDA is a probabilistic algorithm that classifies ED visits into 4 categories including non-emergent and avoidable, emergent but treatable in primary care, emergent but preventable if treated immediately, and emergent care was unavoidable (Ballard et al. 2010). We created a binary indicator to identify potentially avoidable ED visits where the NYU-EDA probabilistic score for non-emergent and avoidable was greater than zero.

After meeting their deductible and moving to a decreased cost-sharing structure, individuals may choose to postpone healthcare that is exempt from cost-sharing (e.g. preventive care) while increasing utilization for healthcare that is subject to cost-sharing (Fronstin, Roebuck, Buxbaum, and Fendrick 2020). To examine this potential mechanism, we conducted an analysis to investigate whether utilization of preventive care increased in the subsequent benefit period among those who met their family deductible in the prior benefit period.

**Results**

*Sample Characteristics*

Our analysis sample included 126,386 unique individuals, of which 7,714 had a family member who experienced an accidental injury (6.1%). Table 1 shows baseline characteristics for the non-injured individuals in our sample grouped by whether they had a family member experience an accidental injury. Accidental injuries were more likely to occur in larger families, likely because larger families have more members who could suffer an accidental injury. Asian populations and individuals aged 51 and older were less likely to have a family member with an accidental injury. The distribution of annual family deductible amount was similar between groups. Appendix Table A3 provides the distribution of accidental injuries grouped by ICD-10 code. Diagnosis codes for injuries to a specific region of the body ("S" codes) accounted for 87% of accidental injuries, with head and arm injuries being the most common. Diagnosis codes related to burn and poisoning injuries were less common, comprising 5% of accidental injuries. Appendix Table A4 provides the distribution of accidental injuries by AIS injury severity index, 73% were classified as minor, 20% as moderate, and 7% as serious or more severe. Baseline characteristics for non-injured individuals by whether their family deductible was met are presented in Appendix Table A5, showing that individuals with lower family deductibles and those with an increased comorbidity burden were more likely to meet their annual deductible.

*First Stage*

Individuals with family members who sustained an accidental injury were more than twice as likely to meet their family deductible as compared to individuals without a family member sustaining an accidental injury. In the first stage, the probability of meeting the family deductible was 14.7 percentage points (pp) larger [95% CI: 14.1 to 15.4 pp] in families with an accidental injury. When estimated as a linear probability model, our first-stage F-Statistic exceeded commonly accepted thresholds for a strong instrument ($F(1,270035)=1781.8$, p-value < 0.001) (Lee, McCrary, Moreira, and Porter 2022). The corresponding first stage Chi-square statistic from our 2SRI logistic regression model was 2768.5 (p-value < 0.001). Comprehensive output from our first stage logistic model is reported in Appendix Table A6.

To explore the exogeneity of our instrument, we examined the distribution of accidental injuries by month of benefit period as shown in Appendix Table A7. We observed a uniform distribution across month of benefit period, providing evidence that accidental injuries were random events that are not correlated with time periods that might have increased utilization (e.g. the last month of a benefit period when utilization may increase). In addition, we conducted a placebo test to investigate whether a family member's accidental injury in the current benefit period predicted meeting the family deductible in the previous benefit period. A priori, we would not expect an accidental injury in the future to influence meeting the family deductible in the past, unless there is an unobserved endogenous factor associated with a family's risk tolerance and healthcare preferences. Findings from our placebo test indicated that a future accidental injury did not predict meeting the family deductible in the past [p-value: 0.573].

*Second Stage Results*

Results from the second stage of our 2SRI models found that meeting a family deductible increased emergency department utilization by 10.0 pp [95% CI: 5.7 to 14.3 pp] (Table 2). Meeting the family deductible also increased avoidable ED visits by 2.4 pp [95% CI: 0.2 to 4.7]. We did not observe statistically significant differences for outpatient clinic visits, inpatient admissions, urgent care, or ambulatory surgery. Among preventive care utilization, we found that meeting a family deductible decreased wellness visits by 5.7 pp [95% CI: -8.9 to -2.5 pp]. Among procedural outcomes, meeting the family deductible decreased venipuncture by 2.7 pp [95% CI: -4.8 to -0.6 pp].

In contrast, naïve logistic regression models showed that meeting a family deductible increased utilization for all outcomes examined, regardless of whether the type of care was subject to the deductible shown in Appendix Table A8. For example, outpatient clinic visits increased by 12.5 pp [95% CI: 11.9 to 13.0 pp], mammography by 1.8 pp [95% CI: 0.1 to 3.5 pp], and venipuncture procedures increased by 3.1 pp [95% CI: 2.8 to 3.5 pp].

*Sensitivity Analyses*

Table 3 presents results from our sensitivity analyses. Models examining heterogeneous effects found individuals 51 and older did not increase ED utilization after meeting the family deductible. The negative effect on wellness visits found in our primary result was not present among individuals aged 31 to 50. When excluding severe accidental injuries, we found a reduction in mammography uptake of 9.6 pp [95% CI: -21.0 to -1.8 pp] that we did not observe in our primary specification. Primary results were robust to different deductible amounts, excluding individuals meeting their out-of-pocket maximum, and excluding December index dates when healthcare access may be restricted. Analyses examining utilization of preventive care in the subsequent benefit period among those who met their family deductible in the prior benefit period showed that wellness visits increased by 3.7 pp in the year after meeting a family deductible.

**Discussion**

In this study, we estimated the causal effect of a reduction in patient cost-sharing after meeting a family deductible on healthcare utilization among people enrolled in family-level health plans with deductibles. We accounted for endogeneity related to health plan selection by using a family member's accidental injury as a pseudo-randomizer that increased the probability the non-injured family member met their family deductible. While many studies have explored the effects of deductibles and cost-sharing on utilization outcomes through employer mandated switches from low cost-sharing to high cost-sharing health plans, or benefit changes following the ACA, our study leveraged a novel dataset that explored how an unanticipated reduction in cost-sharing affected future healthcare utilization. By using data from an integrated healthcare system, we held the quality of healthcare system and the provider network constant. Naïve estimates showed a reduction in cost-sharing led to increased utilization for all examined outcomes. However, our primary 2SRI results showed a reduction in cost-sharing increased utilization of emergency department visits by 10.0 pp, but reduced utilization of wellness visits by 5.7 pp. While other outcomes did not meet statistical significance at the 95% level, the direction of the effect implied a reduced cost-sharing structure may also increase urgent care, outpatient clinic visits, and office visit procedures, while reducing screening utilization such as mammography. The empirical strategy we leveraged produced a LATE, the effect for individuals who changed utilization patterns following a reduction in cost-sharing because they met their family deductible due to a family member's accidental injury.

Our finding that a change to the patient's cost-sharing structure influenced utilization patterns for healthcare that is subject to cost-sharing aligns with a robust literature (Aron-Dine, Einav, and Finkelstein 2013; Agarwal, Mazurenko, and Menachemi 2017; Reddy et al. 2014). For example, prior research has estimated that increasing emergency department copayments by $20 to $35, decreased utilization by 12% (Chandra, Gruber, and McKnight 2010; Hsu et al. 2006). We also found that 24% of the observed increase in ED utilization was potentially avoidable, and may have been treatable in an ambulatory care setting. With an average cost exceeding $3,500 per ED visit, appropriately redirecting non-emergent care to urgent care or primary care settings could lead to substantial cost savings with a minimal impact on health outcomes (Lane, Mallow, Hooker, and Hooker 2020; Weinick, Burns, and Mehrotra 2010). Conversely, 76% of the increased emergency department utilization was potentially necessary, indicating higher cost-sharing obligations may cause individuals to forgo necessary emergency care, which can increase the need for subsequent hospitalizations (Wharam et al. 2013). In contrast to previous studies that showed individuals with lower cost-sharing obligations used ED utilization as a substitute for primary care, we observed ED utilization was being used as a complement to primary care, potentially exacerbating overcrowding issues faced by many ED's (Denham et al. 2024; Begley, Vojvodic, Seo, and Burau 2006). This finding might be attributable to our study setting, in which coordinated healthcare was provided to an insured population. Previous studies that found ED was a substitute for primary care examined Medicaid and uninsured populations, who often received care from non-coordinated providers or lacked a usual place of care (Billings, Parikh, and Mijanovich 2000; Chou, Venkatesh, Trueger, and Pitts 2019; Capp et al. 2017).

We observed that individuals who met their family deductible reduced utilization of wellness visits. This finding is contrary to prior studies, which have reported that reduced cost-sharing either increases or has no effect on care that is exempt from the deductible (Norris et al. 2022). When transitioning to a reduced cost-sharing structure, individuals might become more price-conscious, opting for higher-cost healthcare services (e.g., ED visits) while delaying healthcare not subject to cost-sharing until the next benefit period. Our sensitivity analyses showed an increase in wellness visits in the year after meeting the family deductible, suggesting individuals were postponing utilization for certain types of preventive care not subject to cost-sharing after meeting their deductible.

Sensitivity analyses examining heterogeneous effects showed increased ED utilization was limited to individuals 50 and younger, a finding that agrees with prior studies indicating younger populations may prefer ED care over primary care when out-of-pocket costs are comparable (Uscher-Pines et al. 2013). Interventions tailored at educating patients to utilize more appropriate care, such as primary care and urgent care clinics could alleviate overuse of the emergency department (Llovera et al. 2019). This strategy is promoted by KPCO, which offers services at multiple urgent care locations.

In contrast to previous studies, whether the individual faced a high-deductible or low-deductible did not impact our findings (Borah, Burns, and Shah 2011; Wharam et al. 2018; Beeuwkes Buntin, Haviland, McDevitt, and Sood 2011; Mazurenko, Buntin, and Menachemi 2019; Agarwal, Mazurenko, and Menachemi 2017; Wain et al. 2023). Moreover, excluding individuals meeting their out-of-pocket maximum did not change our findings. While individuals who meet their out-of-pocket maximum are likely to be heavy users of the healthcare system, excluding their utilization did not impact our results. However, meeting the family out-of-pocket maximum was relatively rare in our sample, only 1% of individuals met their out-of-pocket maximum, cohorts with a higher proportion of individuals meeting the out-of-pocket maximum may find different utilization patterns. Excluding those who met their family deductible late in the calendar year did not impact our findings, a time-period when access to healthcare may become more restricted, particularly for ED access (Janke, Melnick, and Venkatesh 2022).

This study has several limitations. The treatment effects presented are interpreted as a LATE for compliers and do not represent an average treatment effect of meeting a family deductible on utilization. In this study, a complier refers to an individual who adjusts their utilization patterns after meeting a family deductible due to an accidental injury sustained by a family member. Thus, the compliers are family members that experience an exogenous shock that forces them to meet their family deductible amount. We were unable to observe whether an individual has access to an HRA or HSA to supplement their initial healthcare costs so we cannot examine findings by the impact of meeting the deductible on family finances. To allow a person to meet their family deductible without meeting their individual deductible, we limited our sample to those enrolled on family plans with 3 or more members, which limits the external validity of our findings. Furthermore, we excluded families with more than one member suffering an injury during a benefit period because they may face different barriers to healthcare consumption. Our results may not extend to those enrolled in individual plans, families with 2 people enrolled, or families experiencing multiple injuries. Our study focused on utilization in primary and emergent care settings, our results may not extend to specialty care utilization which can have unique access barriers (Marsh, Kersel, Havill, and Sleigh 1998). While our 2SRI methods were designed to addresses endogeneity related to health plan selection and meeting a family deductible, it is possible there are unmeasured factors related to a family member's accidental injury that directly predict subsequent utilization of the uninjured family member.

**Conclusion**

Our findings showed that meeting a family deductible increased utilization for emergency department care while decreasing utilization for some types of preventive care. A portion of the increased ED utilization was non-emergent and potentially avoidable. Given the high cost of ED visits, interventions tailored to redirecting non-emergent visits to primary care or urgent care settings among who reach their deductible could result in significant cost savings without harming health outcomes. Conversely, a portion of the increased ED utilization was for high-severity visits, necessary care that might have been neglected if the individual did not meet their family deductible and moved to a decreased cost-sharing structure. Improving appropriate utilization of ED services through a multi-faceted approach, including patient education, prompt triage of patients, and lowered cost-sharing incentives for necessary emergency care utilization could decrease subsequent hospitalizations linked to missed ED utilization for high-severity needs. The decreased preventive care utilization we observed may lead to worse health outcomes and increased down-stream medical costs. Identifying and addressing care gaps through coordinated care efforts, particularly in ED settings, can help improve uptake of preventive care. Future studies can examine utilization outcomes longitudinally over multiple years and explore heterogeneous effects among individuals enrolled in plans with tax-free spending accounts.

**Figure 1** Cost-Sharing in Plans with Deductibles

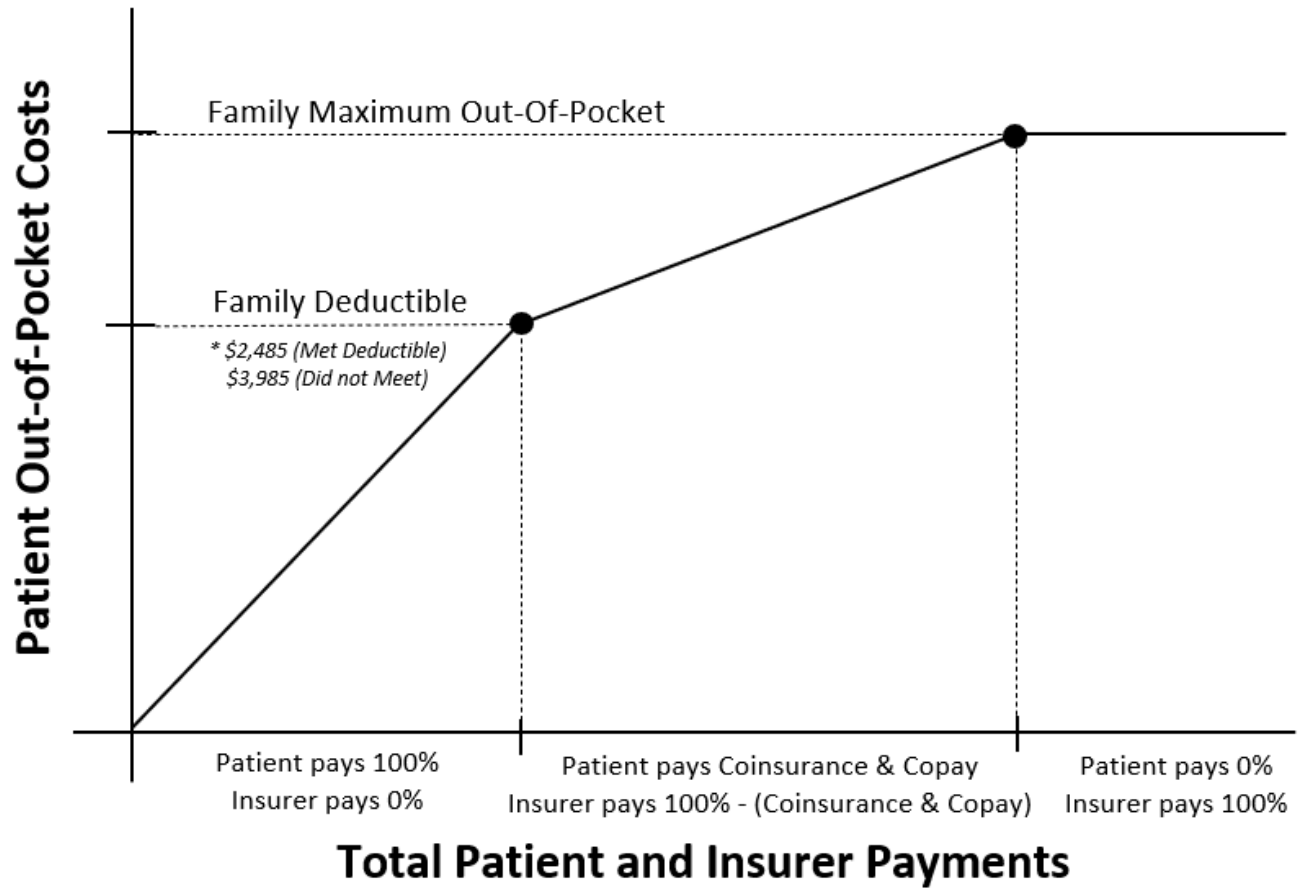

* Mean deductible values for individuals in cohort who met their annual family deductible and did not meet their annual family deductible.

**Figure 2** Instrumental Variable Approach

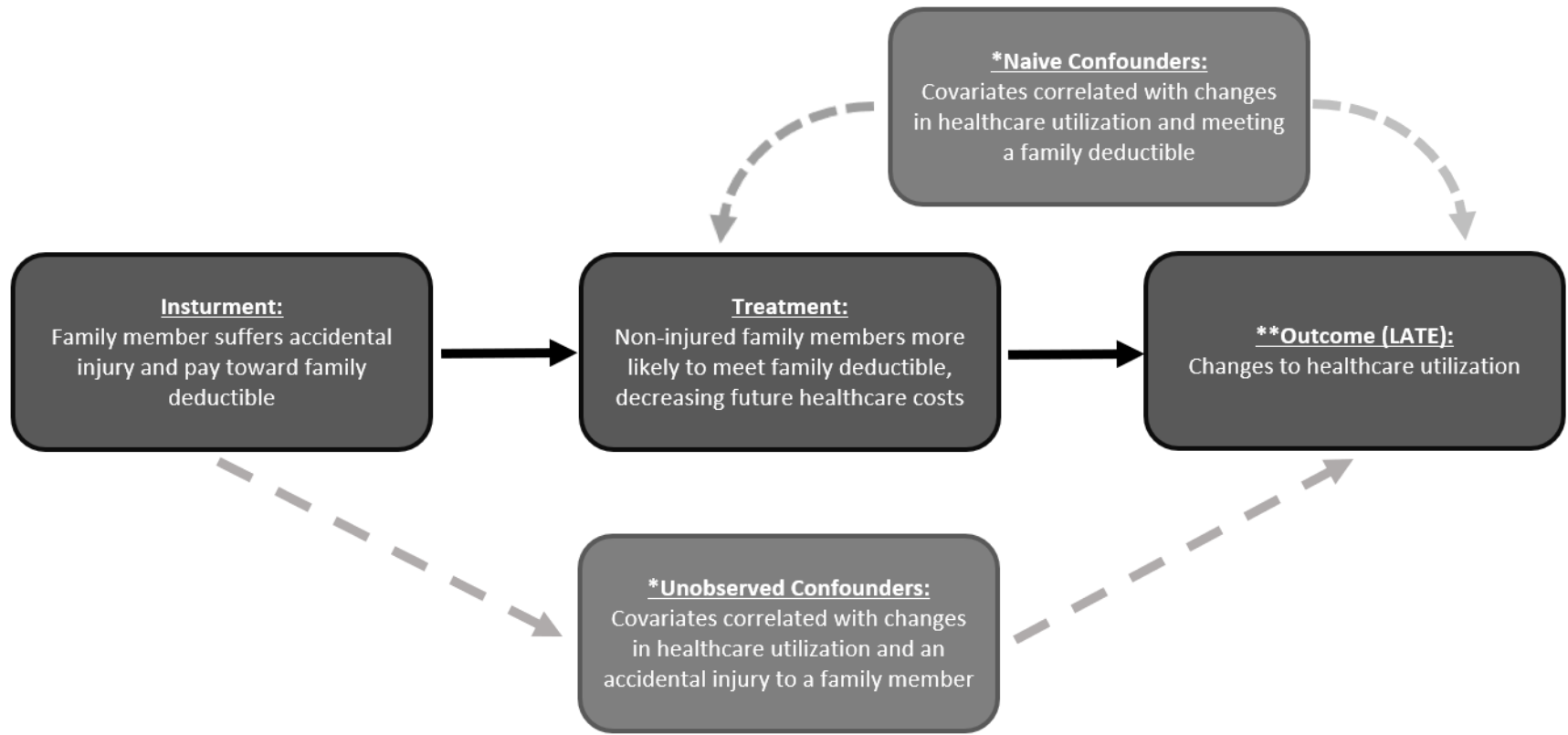

*The treatment variable in a properly specified instrumental variable analysis is independent of both naïve and unobserved confounders.
**Changes in the outcome are interpreted as local average treatment effects (LATE), changes to healthcare utilization after meeting a family deductible due to a family member suffering an accidental injury.

**Table 1** Baseline Characteristics of Non-Injured Individuals by Family Member Experiencing Accidental Injury

|  | No | | Yes | | Total | | p-value |
|---|---|---|---|---|---|---|---|
|  | N | % | N | % | N | % |  |
| Total | 118,672 |  | 7,714 |  | 126,386 |  |  |
| Race/Ethnicity |  |  |  |  |  |  | *<0.001* |
|     Asian | 5,106 | 4.3% | 190 | 2.5% | 5,296 | 4.2% |  |
|     Black | 3,656 | 3.1% | 221 | 2.9% | 3,877 | 3.1% |  |
|     Hispanic | 6,452 | 5.4% | 352 | 4.6% | 6,804 | 5.4% |  |
|     White Non-Hispanic | 70,867 | 59.7% | 4,686 | 60.7% | 75,553 | 59.8% |  |
|     Other | 5,881 | 5.0% | 369 | 4.8% | 6,250 | 4.9% |  |
|     Unknown | 26,710 | 22.5% | 1,896 | 24.6% | 28,606 | 22.6% |  |
| Age at Baseline |  |  |  |  |  |  | *<0.001* |
|     18 to 26 | 36,434 | 30.7% | 2,394 | 31.0% | 38,828 | 30.7% |  |
|     27 to 40 | 31,023 | 26.1% | 1,989 | 25.8% | 33,012 | 26.1% |  |
|     41 to 50 | 31,084 | 26.2% | 2,164 | 28.1% | 33,248 | 26.3% |  |
|     51 to 64 | 20,131 | 17.0% | 1,167 | 15.1% | 21,298 | 16.9% |  |
| Family Deductible Amount |  |  |  |  |  |  | *0.033* |
|     less than $1,000 | 24,914 | 21.0% | 1,575 | 20.4% | 26,489 | 21.0% |  |
|     $1,001 to $2,500 | 30,235 | 25.5% | 2,069 | 26.8% | 32,304 | 25.6% |  |
|     $2,501 to $5,000 | 34,065 | 28.7% | 2,225 | 28.8% | 36,290 | 28.7% |  |
|     $5,000 or more | 29,458 | 24.8% | 1,845 | 23.9% | 31,303 | 24.8% |  |
| Yost Quintile |  |  |  |  |  |  | *0.35* |
|     1 | 8,550 | 7.2% | 562 | 7.3% | 9,112 | 7.2% |  |
|     2 | 13,030 | 11.0% | 798 | 10.3% | 13,828 | 10.9% |  |
|     3 | 20,077 | 16.9% | 1,326 | 17.2% | 21,403 | 16.9% |  |
|     4 | 30,426 | 25.6% | 2,027 | 26.3% | 32,453 | 25.7% |  |
|     5 | 46,589 | 39.3% | 3,001 | 38.9% | 49,590 | 39.2% |  |
| Family Size |  |  |  |  |  |  | *<0.001* |
|     3 | 45,688 | 38.5% | 1,953 | 25.3% | 47,641 | 37.7% |  |
|     4 or more | 72,984 | 61.5% | 5,761 | 74.7% | 78,745 | 62.3% |  |
| Elixhauser Comorbidity Index |  |  |  |  |  |  | *0.14* |
|     0 | 87,201 | 73.5% | 5,588 | 72.4% | 92,789 | 73.4% |  |
|     1 | 19,959 | 16.8% | 1,335 | 17.3% | 21,294 | 16.8% |  |
|     2 | 6,944 | 5.9% | 462 | 6.0% | 7,406 | 5.9% |  |
|     3+ | 4,568 | 3.8% | 329 | 4.3% | 4,897 | 3.9% |  |

Baseline characteristics measured at first index-date per unique individual.
"Other Race" includes Native Hawaiian/Pacific Islanders, American Indian/Alaskan Natives, and multi-race populations.

**Table 2** Second Stage of 2SRI Models: Marginal Effect of Meeting Family Deductible on Utilization of Healthcare Services for All Deductible Levels

| Type of Care | Obs | M.E. | 95% CI |
|---|---|---|---|
| **Utilization by Setting** | | | |
| Emergency Department | 269,919 | 10.0 | [5.7 to 14.3] |
| Inpatient Admissions | 269,919 | 0.3 | [-0.9 to 1.5] |
| Urgent Care | 269,919 | 1.7 | [-0.8 to 4.2] |
| Outpatient Clinic Visits | 269,919 | 2.8 | [-2.3 to 8] |
| Ambulatory Surgery | 269,919 | 0.4 | [-1.2 to 2.1] |
| **Procedures** | | | |
| Office Visits | 269,919 | 3.0 | [-1.2 to 7.2] |
| Venipuncture | 269,919 | -2.7 | [-4.8 to -0.6] |
| Physical Therapy | 269,919 | -0.6 | [-1.6 to 0.5] |
| **Annual Preventive Care** | | | |
| Mammography | 24,171 | -9.4 | [-20.1 to 1.3] |
| Wellness Visit | 232,331 | -5.7 | [-8.9 to -2.5] |

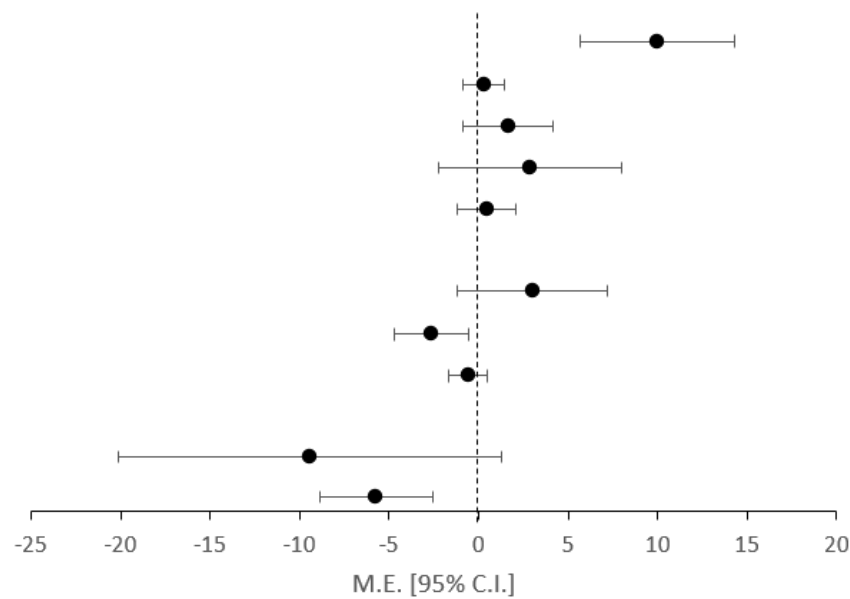

All deductible levels were include in our primary model estimates
Marginal effects (M.E.) were calculated at mean values for all other predictor variables
95% confidence intervals are presented in brackets to the right of the marginal effect estimates
Total observations (Obs) is presented for each model estimate

**Table 3** Sensitivity of Second Stage of 2SRI Models: Marginal Effect of Meeting Family Deductible on Utilization of Healthcare Services Among Different Sub-Samples

| Type of Care | High-Deductible Only | | | Low-Deductible Only | | | Exclude Met Family Out-Of-Pocket Maximum | | |
|---|---|---|---|---|---|---|---|---|---|
| | Obs | M.E. | 95% C.I. | Obs | M.E. | 95% C.I. | Obs | M.E. | 95% C.I. |
| **Utilization by Setting** | | | | | | | | | |
| Emergency Department | 142,629 | 11.9 | [0.5 to 19.3] | 127,290 | 8.0 | [2.5 to 13.5] | 260,975 | 8.2 | [3.4 to 13.1] |
| Inpatient Admissions | 142,629 | -0.2 | [-1.7 to 1.2] | 127,290 | 1.0 | [-1.2 to 3.2] | 260,975 | 0.1 | [-1 to 1.2] |
| Urgent Care | 142,629 | 0.9 | [-2.3 to 4.1] | 127,290 | 2.8 | [-1.1 to 6.7] | 260,975 | 0.7 | [-1.8 to 3.2] |
| Outpatient Clinic Visits | 142,629 | 3.8 | [-3.8 to 11.5] | 127,290 | 2.5 | [-4.4 to 9.5] | 260,975 | -0.1 | [-5.5 to 5.2] |
| Ambulatory Surgery | 142,629 | 1.1 | [-1.3 to 3.5] | 127,290 | 0.0 | [-2.1 to 2.1] | 260,975 | -0.5 | [-1.9 to 0.9] |
| **Procedures** | | | | | | | | | |
| Office Visits | 142,629 | 4.7 | [-1.3 to 10.7] | 127,290 | 1.5 | [-4.7 to 7.7] | 260,975 | 0.2 | [-4.1 to 4.5] |
| Venipuncture | 142,629 | -3.7 | [-6.3 to -1.1] | 127,290 | -0.9 | [-4.3 to 2.5] | 260,975 | -3.2 | [-5.3 to -1.1] |
| Physical Therapy | 142,629 | -0.2 | [-1.7 to 1.3] | 127,290 | -0.9 | [-2.6 to 0.7] | 260,975 | -0.8 | [-2.1 to 0.6] |
| **Annual Preventive Care** | | | | | | | | | |
| Mammography | 13,182 | -6.0 | [-21.3 to 9.4] | 10,989 | -14.5 | [-29.2 to -0.2] | 23,370 | -9.6 | [-21.9 to 2.6] |
| Wellness Visit | 122,258 | -6.6 | [-11.3 to -1.9] | 110,073 | -4.7 | [-9 to -0.5] | 224,406 | -6.5 | [-9.7 to -3.2] |

Marginal effects (M.E.) were calculated at mean values for all other predictor variables.
95% confidence intervals are presented in brackets to the right of the marginal effect estimates.
Total observations (Obs) is presented for each model estimate.

**Table 3 (continued)** Sensitivity of Second Stage of 2SRI Models: Marginal Effect of Meeting Family Deductible on Utilization of Healthcare Services Among Different Sub-Samples

| Type of Care | Limit to Families First Accidental Injury | | | Exclude Severe Injuries | | | Exclude December Index Dates | | |
|---|---|---|---|---|---|---|---|---|---|
| | Obs | M.E. | 95% C.I. | Obs | M.E. | 95% C.I. | Obs | M.E. | 95% C.I. |
| **Utilization by Setting** | | | | | | | | | |
|   Emergency Department | 265,606 | 10.6 | [5.3 to 15.9] | 269,708 | 10.1 | [5.5 to 14.7] | 243,815 | 10.9 | [6 to 15.7] |
|   Inpatient Admissions | 265,606 | 0.6 | [-0.8 to 2] | 269,708 | 0.3 | [-0.9 to 1.4] | 243,815 | 0.3 | [-0.9 to 1.4] |
|   Urgent Care | 265,606 | 1.2 | [-1.5 to 3.9] | 269,708 | 1.9 | [-0.6 to 4.3] | 243,815 | 1.6 | [-1.1 to 4.3] |
|   Outpatient Clinic Visits | 265,606 | 2.9 | [-2.6 to 8.5] | 269,708 | 2.7 | [-2.1 to 7.5] | 243,815 | 2.1 | [-3.1 to 7.4] |
|   Ambulatory Surgery | 265,606 | 0.2 | [-1.4 to 1.9] | 269,708 | 0.4 | [-1.1 to 2] | 243,815 | 0.4 | [-1.2 to 2.1] |
| **Procedures** | | | | | | | | | |
|   Office Visits | 265,606 | 2.9 | [-1.9 to 7.6] | 269,708 | 3.1 | [-1.2 to 7.3] | 243,815 | 2.7 | [-1.9 to 7.4] |
|   Venipuncture | 265,606 | -2.4 | [-4.8 to -0.1] | 269,708 | -2.6 | [-4.7 to -0.5] | 243,815 | -2.6 | [-4.9 to -0.3] |
|   Physical Therapy | 265,606 | -0.4 | [-1.7 to 0.8] | 269,708 | -0.5 | [-1.6 to 0.5] | 243,815 | -0.7 | [-1.8 to 0.4] |
| **Annual Preventive Care** | | | | | | | | | |
|   Mammography | 23,800 | -10.9 | [-22.8 to 1] | 24,147 | -9.6 | [-21 to -1.8] | 22,200 | -10.9 | [-22.5 to 0.6] |
|   Wellness Visit | 228,607 | -6.0 | [-9.7 to -2.7] | 232,149 | -5.9 | [-9.2 to -2.6] | 212,399 | -6.1 | [-9.4 to -2.7] |

Marginal effects (M.E.) were calculated at mean values for all other predictor variables.
95% confidence intervals are presented in brackets to the right of the marginal effect estimates.
Total observations (Obs) is presented for each model estimate.

**Table 3 (continued)** Sensitivity of Second Stage of 2SRI Models: Marginal Effect of Meeting Family Deductible on Utilization of Healthcare Services Among Different Sub-Samples

|  | Age <= 30 | | | Age 31 to 50 | | | Age > 50 | | |
|---|---|---|---|---|---|---|---|---|---|
| Type of Care | Obs | M.E. | 95% C.I. | Obs | M.E. | 95% C.I. | Obs | M.E. | 95% C.I. |
| *Utilization by Setting* | | | | | | | | | |
|   Emergency Department | 82,776 | 15.3 | [5 to 25.6] | 132,057 | 9.1 | [2.8 to 15.4] | 51,807 | 2.6 | [-2.7 to 8] |
|   Inpatient Admissions | 82,776 | 0.9 | [-2.4 to 4.1] | 132,057 | 0.5 | [-1.4 to 2.3] | 51,807 | -0.2 | [-3.1 to 2.6] |
|   Urgent Care | 82,776 | 3.7 | [-1.7 to 9.1] | 132,057 | 3.1 | [-1 to 7.2] | 51,807 | -2.3 | [-6 to 1.5] |
|   Outpatient Clinic Visits | 82,776 | 0.0 | [-8.5 to 8.4] | 132,057 | 6.7 | [-0.5 to 14] | 51,807 | -1.0 | [-11.6 to 9.6] |
|   Ambulatory Surgery | 82,776 | 0.4 | [-2 to 2.8] | 132,057 | 0.2 | [-2 to 2.4] | 51,807 | 1.0 | [-4.1 to 6] |
| *Procedures* | | | | | | | | | |
|   Office Visits | 82,776 | 1.9 | [-4.7 to 8.5] | 132,057 | 7.8 | [1.5 to 14.2] | 51,807 | -6.9 | [-14.9 to 1] |
|   Venipuncture | 82,776 | -1.4 | [-4.4 to 1.6] | 132,057 | -2.2 | [-5.4 to 1] | 51,807 | -6.3 | [-11 to -1.7] |
|   Physical Therapy | 82,776 | -0.1 | [-2.2 to 2] | 132,057 | -0.8 | [-2.3 to 7.8] | 51,807 | -0.5 | [-3.9 to 2.8] |
| *Annual Preventive Care* | | | | | | | | | |
|   Mammography | | | N/A | | | N/A | | | N/A |
|   Wellness Visit | 74,372 | -4.2 | [-8.8 to -0.3] | 112,369 | -4.1 | [-9.2 to 1.1] | 42,736 | -9.5 | [-17 to -2] |

Marginal effects (M.E.) were calculated at mean values for all other predictor variables.
95% confidence intervals are presented in brackets to the right of the marginal effect estimates.
Total observations (Obs) is presented for each model estimate.
Mammography eligibility is limited to women aged 50 to 74, thus it is not possible to examine outcomes stratified by gender or age categories.

**Table 3 (continued)** Sensitivity of Second Stage of 2SRI Models: Marginal Effect of Meeting Family Deductible on Utilization of Healthcare Among Different Sub-Samples

|  | Female Only | | | Male Only | | |
| --- | --- | --- | --- | --- | --- | --- |
| Type of Care | Obs | M.E. | 95% C.I. | Obs | M.E. | 95% C.I. |
| *Utilization by Setting* | | | | | | |
|   Emergency Department | 137,910 | 8.0 | [2.7 to 13.3] | 131,939 | 13.9 | [5.5 to 22.3] |
|   Inpatient Admissions | 137,910 | -0.2 | [-1.8 to 1.5] | 131,939 | 0.7 | [-1.4 to 2.8] |
|   Urgent Care | 137,910 | 1.2 | [-2.3 to 4.7] | 131,939 | 2.1 | [-1.4 to 5.6] |
|   Outpatient Clinic Visits | 137,910 | 4.5 | [-2.3 to 11.2] | 131,939 | 1.2 | [-5.5 to 7.8] |
|   Ambulatory Surgery | 137,910 | 0.0 | [-2.2 to 2.1] | 131,939 | 1.0 | [-1.5 to 3.5] |
| *Procedures* | | | | | | |
|   Office Visits | 137,910 | 4.3 | [-1.7 to 10.4] | 131,939 | 1.6 | [-3.9 to 7.2] |
|   Venipuncture | 137,910 | -2.7 | [-5.8 to 0.4] | 131,939 | -2.8 | [-5.5 to -0.1] |
|   Physical Therapy | 137,910 | -0.3 | [-1.9 to 1.4] | 131,939 | -1.0 | [-2.8 to 0.8] |
| *Annual Preventive Care* | | | | | | |
|   Mammography | N/A | | N/A | | | N/A |
|   Wellness Visit | 114,247 | -6.5 | [-11.7 to -1.4] | 118,021 | -5.4 | [-8.9 to -1.9] |

Marginal effects (M.E.) were calculated at mean values for all other predictor variables.
95% confidence intervals are presented in brackets to the right of the marginal effect estimates.
Total observations (Obs) is presented for each model estimate.
Mammography eligibility is limited to women aged 50 to 74, thus it is not possible to examine outcomes stratified by gender or age categories.

# APPENDIX:

**Table A1** Code List for Outcome Variables

| Outcome | Description | Code Type | Code |
|---|---|---|---|
| Wellness Visit | 18-39: initial periodic comprehensive preventive medicine | CPT | 99385 |
| | 40-64: initial periodic comprehensive preventive medicine | CPT | 99386 |
| | 18-39: subsequent periodic comprehensive preventive medicine | CPT | 99395 |
| | 40-64: subsequent periodic comprehensive preventive medicine | CPT | 99396 |
| Mammography | computer aided detection screening mammogram | CPT | 76083 |
| | mammogram, one breast x-ray breast e-c | CPT | 76090 |
| | mammogram, both breasts x-ray breast e-c | CPT | 76091 |
| | mammogram, screening x-ray breast e-c | CPT | 76092 |
| | screening mammography | CPT | 77052 |
| | screening mammography, bilateral | CPT | 77057 |
| | screening mammography, bilateral | CPT | 77067 |
| | screening mammography, bilateral | HCPCS | G0202 |

The "CPT" code type represents the Current Procedural Terminology code set developed by the American Medical Association

The "HCPCS" code type represents Healthcare Common Procedure Coding System developed by the Centers for Medicare and Medicaid Services

**Table A2** ICD-10 Code List for Accidental Injuries

| Diagnostic Category | ICD-10 Code (first 3 of code) | Description |
| --- | --- | --- |
| Injuries to the head | S00 | Superficial injury of head |
| | S01 | Open wound of head |
| | S02 | Fracture of skull and facial bones |
| | S03 | Dislocation and sprain of joints and ligaments of head |
| | S04 | Injury of cranial nerve |
| | S05 | Injury of eye and orbit |
| | S06 | Intracranial injury |
| | S07 | Crushing injury of head |
| | S08 | Avulsion and traumatic amputation of part of head |
| | S09 | Other and unspecified injuries of head |
| Injuries to the neck | S10 | Superficial injury of neck |
| | S11 | Open wound of neck |
| | S12 | Fracture of cervical vertebra and other parts of neck |
| | S13 | Dislocation and sprain of joints and ligaments at neck level |
| | S14 | Injury of nerves and spinal cord at neck level |
| | S15 | Injury of blood vessels at neck level |
| | S16 | Injury of muscle, fascia and tendon at neck level |
| | S17 | Crushing injury of neck |
| | S19 | Other specified and unspecified injuries of neck |
| Injuries to the thorax | S20 | Superficial injury of thorax |
| | S21 | Open wound of thorax |
| | S22 | Fracture of rib(s), sternum and thoracic spine |
| | S23 | Dislocation and sprain of joints and ligaments of thorax |
| | S24 | Injury of nerves and spinal cord at thorax level |
| | S25 | Injury of blood vessels of thorax |
| | S26 | Injury of heart |
| | S27 | Injury of other and unspecified intrathoracic organs |
| | S28 | Crushing injury of thorax, and traumatic amputation of part of thorax |
| | S29 | Other and unspecified injuries of thorax |
| Injuries to the abdomen, lower back, lumbar spine, pelvis and external genitals | S30 | Superficial injury of abdomen, lower back, pelvis and external genitals |
| | S31 | Open wound of abdomen, lower back, pelvis and external genitals |
| | S32 | Fracture of lumbar spine and pelvis |
|  | S33 | Dislocation and sprain of joints and ligaments of lumbar spine and pelvis |
|  | S34 | Injury of lumbar and sacral spinal cord and nerves at abdomen, lower back and pelvis level |
|  | S35 | Injury of blood vessels at abdomen, lower back and pelvis level |
|  | S36 | Injury of intra-abdominal organs |
|  | S37 | Injury of urinary and pelvic organs |
|  | S38 | Crushing injury and traumatic amputation of abdomen, lower back, pelvis and external genitals |
|  | S39 | Other and unspecified injuries of abdomen, lower back, pelvis and external genitals |
| Injuries to the shoulder and upper arm | S40 | Superficial injury of shoulder and upper arm |
|  | S41 | Open wound of shoulder and upper arm |
|  | S42 | Fracture of shoulder and upper arm |
|  | S43 | Dislocation and sprain of joints and ligaments of shoulder girdle |
|  | S44 | Injury of nerves at shoulder and upper arm level |
|  | S45 | Injury of blood vessels at shoulder and upper arm level |
|  | S46 | Injury of muscle, fascia and tendon at shoulder and upper arm level |
|  | S47 | Crushing injury of shoulder and upper arm |
|  | S48 | Traumatic amputation of shoulder and upper arm |
|  | S49 | Other and unspecified injuries of shoulder and upper arm |
| Injuries to the elbow and forearm | S50 | Superficial injury of elbow and forearm |
|  | S51 | Open wound of elbow and forearm |
|  | S52 | Fracture of forearm |
|  | S53 | Dislocation and sprain of joints and ligaments of elbow |
|  | S54 | Injury of nerves at forearm level |
|  | S55 | Injury of blood vessels at forearm level |
|  | S56 | Injury of muscle, fascia and tendon at forearm level |
|  | S57 | Crushing injury of elbow and forearm |
|  | S58 | Traumatic amputation of elbow and forearm |
|  | S59 | Other and unspecified injuries of elbow and forearm |
| Injuries to the wrist, hand and fingers | S60 | Superficial injury of wrist, hand and fingers |
|  | S61 | Open wound of wrist, hand and fingers |
|  | S62 | Fracture at wrist and hand level |
|  | S63 | Dislocation and sprain of joints and ligaments at wrist and hand level |
|  | S64 | Injury of nerves at wrist and hand level |
|  | S65 | Injury of blood vessels at wrist and hand level |
|  | S66 | Injury of muscle, fascia and tendon at wrist and hand level |
|  | S67 | Crushing injury of wrist, hand and fingers |
|  | S68 | Traumatic amputation of wrist, hand and fingers |
|  | S69 | Other and unspecified injuries of wrist, hand and finger(s) |
| Injuries to the hip and thigh | S70 | Superficial injury of hip and thigh |
|  | S71 | Open wound of hip and thigh |
|  | S72 | Fracture of femur |
|  | S73 | Dislocation and sprain of joint and ligaments of hip |
|  | S74 | Injury of nerves at hip and thigh level |
|  | S75 | Injury of blood vessels at hip and thigh level |
|  | S76 | Injury of muscle, fascia and tendon at hip and thigh level |
|  | S77 | Crushing injury of hip and thigh |
|  | S78 | Traumatic amputation of hip and thigh |
|  | S79 | Other and unspecified injuries of hip and thigh |
| Injuries to the knee and lower leg | S80 | Superficial injury of knee and lower leg |
|  | S81 | Open wound of knee and lower leg |
|  | S82 | Fracture of lower leg, including ankle |
|  | S83 | Dislocation and sprain of joints and ligaments of knee |
|  | S84 | Injury of nerves at lower leg level |
|  | S85 | Injury of blood vessels at lower leg level |
|  | S86 | Injury of muscle, fascia and tendon at lower leg level |
|  | S87 | Crushing injury of lower leg |
|  | S88 | Traumatic amputation of lower leg |
|  | S89 | Other and unspecified injuries of lower leg |
| Injuries to the ankle and foot | S90 | Superficial injury of ankle, foot and toes |
|  | S91 | Open wound of ankle, foot and toes |
|  | S92 | Fracture of foot and toe, except ankle |
|  | S93 | Dislocation and sprain of joints and ligaments at ankle, foot and toe level |
|  | S94 | Injury of nerves at ankle and foot level |
|  | S95 | Injury of blood vessels at ankle and foot level |
|  | S96 | Injury of muscle and tendon at ankle and foot level |
|  | S97 | Crushing injury of ankle and foot |
|  | S98 | Traumatic amputation of ankle and foot |
|  | S99 | Other and unspecified injuries of ankle and foot |
| Injuries involving multiple body regions | T07 | Injuries involving multiple body regions |
| Injury of unspecified body region | T14 | Injury of unspecified body region |
| Effects of foreign body entering through natural orifice | T15 | Foreign body on external eye |
| | T16 | Foreign body in ear |
| | T17 | Foreign body in respiratory tract |
| | T18 | Foreign body in alimentary tract |
| | T19 | Foreign body in genitourinary tract |
| Burns and corrosions of external body surface, specified by site | T20 | Burn and corrosion of head, face, and neck |
| | T21 | Burn and corrosion of trunk |
| | T22 | Burn and corrosion of shoulder and upper limb, except wrist and hand |
| | T23 | Burn and corrosion of wrist and hand |
| | T24 | Burn and corrosion of lower limb, except ankle and foot |
| | T25 | Burn and corrosion of ankle and foot |
| Burns and corrosions confined to eye and internal organs | T26 | Burn and corrosion confined to eye and adnexa |
| | T27 | Burn and corrosion of respiratory tract |
| | T28 | Burn and corrosion of other internal organs |
| Burns and corrosions of multiple and unspecified body regions | T30 | Burn and corrosion, body region unspecified |
| | T31 | Burns classified according to extent of body surface involved |
| | T32 | Corrosions classified according to extent of body surface involved |
| Frostbite | T34 | Frostbite with tissue necrosis |
| Toxic effects of substances chiefly nonmedicinal as to source | T52 | Toxic effect of organic solvents |
| | T53 | Toxic effect of halogen derivatives of aliphatic and aromatic hydrocarbons |
| | T54 | Toxic effect of corrosive substances |
| | T55 | Toxic effect of soaps and detergents |
| | T56 | Toxic effect of metals |
| | T57 | Toxic effect of other inorganic substances |
| | T58 | Toxic effect of carbon monoxide |
| | T59 | Toxic effect of other gases, fumes and vapors |
| | T60 | Toxic effect of pesticides |
| | T61 | Toxic effect of noxious substances eaten as seafood |
| | T62 | Toxic effect of other noxious substances eaten as food |
| | T63 | Toxic effect of contact with venomous animals and plants |
| | T64 | Toxic effect of aflatoxin and other mycotoxin food contaminants |
| | T65 | Toxic effect of other and unspecified substances |
| Other and unspecified effects of external causes | T66 | Radiation sickness, unspecified |
| | T67 | Effects of heat and light |
| | T68 | Hypothermia |
| | T69 | Other effects of reduced temperature |
| | T70 | Effects of air pressure and water pressure |
| Certain early complications of trauma | T79 | Certain early complications of trauma |

ICD-10 S and T codes for injuries that were deemed non-accidental were excluded
Only the first 3 alpha-numeric for each ICD-10 code is presented in this table, a comprehensive list of ICD10 codes is available upon request
Groupings for injuries represent those developed by the World Health Organization and the Centers for Disease Control and Management

**Table A3** Frequency of Accidental Injury Grouped by First 3 Characters of ICD-10 Code

| ICD Code Group | Description | Frequency | Percent |
|---|---|---|---|
| S00-S09 | Injuries to the head | 1,944 | 22.5% |
| S10-S19 | Injuries to the neck | 374 | 4.3% |
| S20-S29 | Injuries to the thorax | 239 | 2.8% |
| S30-S39 | Injuries to the abdomen, lower back, lumbar spine, pelvis and external genitals | 368 | 4.3% |
| S40-S49 | Injuries to the shoulder and upper arm | 552 | 6.4% |
| S50-S59 | Injuries to the elbow and forearm | 788 | 9.1% |
| S60-S69 | Injuries to the wrist, hand and fingers | 1,407 | 16.3% |
| S70-S79 | Injuries to the hip and thigh | 190 | 2.2% |
| S80-S89 | Injuries to the knee and lower leg | 849 | 9.8% |
| S90-S99 | Injuries to the ankle and foot | 775 | 9.0% |
| T07-T07 | Injuries involving multiple body regions | 19 | 0.2% |
| T14-T14 | Injury of unspecified body region | 208 | 2.4% |
| T15-T19 | Effects of foreign body entering through natural orifice | 258 | 3.0% |
| T20-T25 | Burns and corrosions of external body surface, specified by site | 37 | 0.4% |
| T26-T28 | Burns and corrosions confined to eye and internal organs | 3 | 0.0% |
| T30-T32 | Burns and corrosions of multiple and unspecified body regions | 50 | 0.6% |
| T34 | Frostbite | 1 | 0.0% |
| T51-T65 | Toxic effects of substances chiefly nonmedicinal as to source | 107 | 1.2% |
| T66-T78 | Other and unspecified effects of external causes | 422 | 4.9% |
| T79-T79 | Certain early complications of trauma | 35 | 0.4% |
| **Total** | **Total** | **8,626** | **100.0%** |

ICD-10 S and T codes for injuries that were deemed non-accidental were excluded

Only the first 3 alpha-numeric for each ICD-10 code is presented in this table, a comprehensive list of ICD10 codes is available upon request

Groupings for injuries represent those developed by the World Health Organization and the Centers for Disease Control and Management

**Table A4** Frequency of Abbreviated Injury Scale (AIS) Classification by Unique Injury

| AIS Score | Frequency | Percent |
|---|---|---|
| 1 - Minor | 6,290 | 72.9% |
| 2 - Moderate | 1,718 | 19.9% |
| 3 - Serious | 539 | 6.2% |
| 4 - Severe | 50 | 0.6% |
| 5 - Critical | 29 | 0.3% |
| 6 - Maximal (Untreatable) | 0 | 0.0% |
| **Total** | **8,626** | **100.0%** |

The AIS algorithm is an anatomically-based injury severity scoring system that classifies each injury by body region
The AIS was developed to score individual patient injuries and their severity

**Table A5** Baseline Characteristics of Non-Injured Individuals by Meeting Family Deductible

|  | No | | Yes | | Total | | p-value |
|---|---|---|---|---|---|---|---|
|  | N | % | N | % | N | % |  |
| Total | 110,681 |  | 15,705 |  | 126,386 |  |  |
| Race/Ethnicity |  |  |  |  |  |  | *<0.001* |
|     Asian | 4,812 | 4.3% | 484 | 3.1% | 5,296 | 4.2% |  |
|     Black | 3,295 | 3.0% | 582 | 3.7% | 3,877 | 3.1% |  |
|     Hispanic | 5,921 | 5.3% | 883 | 5.6% | 6,804 | 5.4% |  |
|     White Non-Hispanic | 66,010 | 59.6% | 9,543 | 60.8% | 75,553 | 59.8% |  |
|     Other | 5,600 | 5.1% | 650 | 4.1% | 6,250 | 4.9% |  |
|     Unknown | 25,043 | 22.6% | 3,563 | 22.7% | 28,606 | 22.6% |  |
| Age at Baseline |  |  |  |  |  |  | *<0.001* |
|     18 to 26 | 33,594 | 30.4% | 5,234 | 33.3% | 38,828 | 30.7% |  |
|     27 to 40 | 28,988 | 26.2% | 4,024 | 25.6% | 33,012 | 26.1% |  |
|     41 to 50 | 29,421 | 26.6% | 3,827 | 24.4% | 33,248 | 26.3% |  |
|     51 to 64 | 18,678 | 16.9% | 2,620 | 16.7% | 21,298 | 16.9% |  |
| Family Deductible Amount |  |  |  |  |  |  | *<0.001* |
|     less than $1,000 | 21,685 | 19.6% | 4,804 | 30.6% | 26,489 | 21.0% |  |
|     $1,001 to $2,500 | 29,423 | 26.6% | 2,881 | 18.3% | 32,304 | 25.6% |  |
|     $2,501 to $5,000 | 29,694 | 26.8% | 6,596 | 42.0% | 36,290 | 28.7% |  |
|     $5,000 or more | 29,879 | 27.0% | 1,424 | 9.1% | 31,303 | 24.8% |  |
| Yost Quintile |  |  |  |  |  |  | *<0.001* |
|     1 | 7,953 | 7.2% | 1,159 | 7.4% | 9,112 | 7.2% |  |
|     2 | 12,187 | 11.0% | 1,641 | 10.4% | 13,828 | 10.9% |  |
|     3 | 18,871 | 17.0% | 2,532 | 16.1% | 21,403 | 16.9% |  |
|     4 | 28,553 | 25.8% | 3,900 | 24.8% | 32,453 | 25.7% |  |
|     5 | 43,117 | 39.0% | 6,473 | 41.2% | 49,590 | 39.2% |  |
| Family Size |  |  |  |  |  |  | *<0.001* |
|     3 | 43,151 | 39.0% | 4,490 | 28.6% | 47,641 | 37.7% |  |
|     4 or more | 67,530 | 61.0% | 11,215 | 71.4% | 78,745 | 62.3% |  |
| Elixhauser Comorbidity Index |  |  |  |  |  |  | *<0.001* |
|     0 | 82,655 | 74.7% | 10,134 | 64.5% | 92,789 | 73.4% |  |
|     1 | 18,145 | 16.4% | 3,149 | 20.1% | 21,294 | 16.8% |  |
|     2 | 6,106 | 5.5% | 1,300 | 8.3% | 7,406 | 5.9% |  |
|     3+ | 3,775 | 3.4% | 1,122 | 7.1% | 4,897 | 3.9% |  |

Baseline characteristics measured at first index-date per unique individual.
"Other Race" includes Native Hawaiian/Pacific Islanders, American Indian/Alaskan Natives, and multi-race populations.

**Table A6** First-Stage Logistic Regression: Parameter Estimates and Standard Errors by Sample Modeled for Effect of Accidental Family Injury on Meeting Family Deductible

| Explanatory Variable | Modeled Sample | | | | | | | | | | |
|---|---|---|---|---|---|---|---|---|---|---|---|
| | All Deductible Levels | High Deductible Only | Low Deductible Only | Exclude Met Family Out-Of-Pocket Max | Exclude Severe Accidental Injuries | Exclude December Index Dates | Age <= 30 years | Age Between 31 and 50 years | Age > 50 years | Female Only | Male Only |
| Family Accidental Injury | 0.968*** | 0.897*** | 1.038*** | 1.008*** | 0.968*** | 0.955*** | 0.934*** | 0.967*** | 1.037*** | 0.974*** | 0.964*** |
| | (0.0192) | (0.0274) | (0.0269) | (0.0210) | (0.0193) | (0.0202) | (0.0343) | (0.0271) | (0.0465) | (0.0268) | (0.0275) |
| No Family Accidental Injury (ref) | | | | | | | | | | | |
| Black Non-Hispanic | 0.144*** | 0.274*** | 0.0166 | 0.127*** | 0.146*** | 0.136*** | 0.208*** | 0.143*** | 0.0357 | 0.118*** | 0.162*** |
| | (0.0313) | (0.0445) | (0.0444) | (0.0345) | (0.0313) | (0.0330) | (0.0528) | (0.0464) | (0.0736) | (0.0451) | (0.0436) |
| Asian | -0.281*** | -0.308*** | -0.252*** | -0.305*** | -0.279*** | -0.297*** | -0.322*** | -0.265*** | -0.305*** | -0.262*** | -0.302*** |
| | (0.0324) | (0.0438) | (0.0481) | (0.0363) | (0.0324) | (0.0340) | (0.0619) | (0.0457) | (0.0715) | (0.0427) | (0.0498) |
| Hispanic | 0.0337* | 0.0983*** | -0.0370 | 0.0675*** | 0.0340* | 0.0239 | 0.0601** | 0.0373 | -0.0275 | 0.0121 | 0.0515* |
| | (0.0186) | (0.0258) | (0.0267) | (0.0203) | (0.0186) | (0.0195) | (0.0293) | (0.0283) | (0.0479) | (0.0261) | (0.0266) |
| Other Race | -0.109*** | -0.118*** | -0.102** | -0.0964*** | -0.110*** | -0.117*** | -0.0752 | -0.117*** | -0.155** | -0.117*** | -0.108*** |
| | (0.0295) | (0.0404) | (0.0430) | (0.0327) | (0.0295) | (0.0309) | (0.0485) | (0.0451) | (0.0673) | (0.0420) | (0.0414) |
| Unknown Race | -0.0438*** | -0.0877*** | 0.00255 | -0.0406** | -0.0434*** | -0.0493*** | -0.0339 | -0.0706*** | 0.0136 | -0.0120 | -0.0725*** |
| | (0.0167) | (0.0232) | (0.0239) | (0.0184) | (0.0167) | (0.0175) | (0.0264) | (0.0256) | (0.0419) | (0.0245) | (0.0229) |
| White (ref) | | | | | | | | | | | |
| Aged 27 to 40 | -0.158*** | -0.120*** | -0.195*** | -0.162*** | -0.158*** | -0.149*** | | | | -0.225*** | -0.0814*** |
| | (0.0161) | (0.0226) | (0.0231) | (0.0178) | (0.0161) | (0.0169) | | | | (0.0219) | (0.0239) |
| Aged 41 to 50 | -0.206*** | -0.144*** | -0.271*** | -0.195*** | -0.207*** | -0.197*** | | | | -0.257*** | -0.153*** |
| | (0.0156) | (0.0217) | (0.0225) | (0.0173) | (0.0156) | (0.0163) | | | | (0.0218) | (0.0224) |
| Aged 51 to 64 | -0.109*** | -0.109*** | -0.108*** | -0.0720*** | -0.108*** | -0.110*** | | | | -0.178*** | -0.0424* |
| | (0.0177) | (0.0249) | (0.0252) | (0.0195) | (0.0177) | (0.0186) | | | | (0.0260) | (0.0244) |
| Aged 18 to 26 (ref) | | | | | | | | | | | |
| Elixhauser Index = 1 | 0.360*** | 0.372*** | 0.348*** | 0.340*** | 0.361*** | 0.360*** | 0.364*** | 0.386*** | 0.297*** | 0.380*** | 0.336*** |
| | (0.0148) | (0.0208) | (0.0210) | (0.0163) | (0.0148) | (0.0155) | (0.0266) | (0.0212) | (0.0338) | (0.0198) | (0.0224) |
| Elixhauser Index = 2 | 0.585*** | 0.598*** | 0.572*** | 0.531*** | 0.586*** | 0.605*** | 0.616*** | 0.603*** | 0.532*** | 0.625*** | 0.533*** |
| | (0.0215) | (0.0310) | (0.0298) | (0.0238) | (0.0215) | (0.0224) | (0.0456) | (0.0302) | (0.0424) | (0.0283) | (0.0333) |
| Elixhauser Index = 3 or more | 0.873*** | 0.946*** | 0.814*** | 0.742*** | 0.873*** | 0.894*** | 1.058*** | 0.840*** | 0.853*** | 0.906*** | 0.839*** |
| | (0.0246) | (0.0363) | (0.0337) | (0.0279) | (0.0246) | (0.0257) | (0.0625) | (0.0357) | (0.0418) | (0.0337) | (0.0361) |
| Elixhauser Index = 0 (ref) | | | | | | | | | | | |
| Family Deductible $1,000 to $2,500 | -0.936*** | | -0.940*** | -1.026*** | -0.937*** | -0.956*** | -0.947*** | -0.915*** | -0.985*** | -0.980*** | -0.888*** |
| | (0.0176) | | (0.0177) | (0.0188) | (0.0176) | (0.0185) | (0.0301) | (0.0262) | (0.0405) | (0.0245) | (0.0253) |
| Family Deductible $2,501 to $5,000 | 0.104*** | 1.502*** | | -0.115*** | 0.104*** | 0.113*** | 0.0310 | 0.192*** | 0.00206 | 0.0702*** | 0.141*** |
| | (0.0139) | (0.0201) | | (0.0150) | (0.0139) | (0.0145) | (0.0245) | (0.0201) | (0.0323) | (0.0194) | (0.0200) |
| Family Deductible $5,001 or more | -1.397*** | | | -1.900*** | -1.395*** | -1.421*** | -1.401*** | -1.402*** | -1.384*** | -1.401*** | -1.392*** |
| | (0.0206) | | | (0.0257) | (0.0206) | (0.0217) | (0.0372) | (0.0305) | (0.0435) | (0.0287) | (0.0295) |
| Family Deductible LT $1,000 (ref) | | | | | | | | | | | |
| Family Count = 4 or more | 0.478*** | 0.518*** | 0.434*** | 0.468*** | 0.477*** | 0.488*** | 0.559*** | 0.425*** | 0.535*** | 0.464*** | 0.491*** |
| | (0.0131) | (0.0184) | (0.0188) | (0.0145) | (0.0131) | (0.0138) | (0.0237) | (0.0197) | (0.0276) | (0.0182) | (0.0191) |
| Family Count = 3 (ref) | | | | | | | | | | | |
| Yost Index | 0.0520*** | 0.0762*** | 0.0264*** | 0.0491*** | 0.0523*** | 0.0528*** | 0.0517*** | 0.0397*** | 0.0828*** | 0.0499*** | 0.0541*** |
| | (0.00490) | (0.00689) | (0.00699) | (0.00540) | (0.00490) | (0.00513) | (0.00841) | (0.00717) | (0.0116) | (0.00683) | (0.00704) |
| Constant | -2.108*** | -3.659*** | -1.946*** | -2.167*** | -2.109*** | -2.102*** | -2.151*** | -2.266*** | -2.303*** | -2.040*** | -2.178*** |
| | (0.0266) | (0.0394) | (0.0364) | (0.0291) | (0.0266) | (0.0279) | (0.0439) | (0.0382) | (0.0579) | (0.0372) | (0.0381) |
| Observations | 269,919 | 142,629 | 127,290 | 260,975 | 269,708 | 243,815 | 82,776 | 132,057 | 51,807 | 137,910 | 131,939 |

Standard errors in parentheses *** p<0.01, ** p<0.05, * p<0.1

We report detailed information on parameter estimates for each outcome in our primary analysis. Parameter estimates are reported as odds ratios from logistic regression models. Asterisks identify explanatory variables that are statistically significant, one asterisk is significant at the 0.10 percent level, two asterisks is significant at the 0.05 percent level, and three asterisks is significant at the 0.01 percent level.

**Table A7** Distribution of Accidental Injuries by Month of Benefit Period

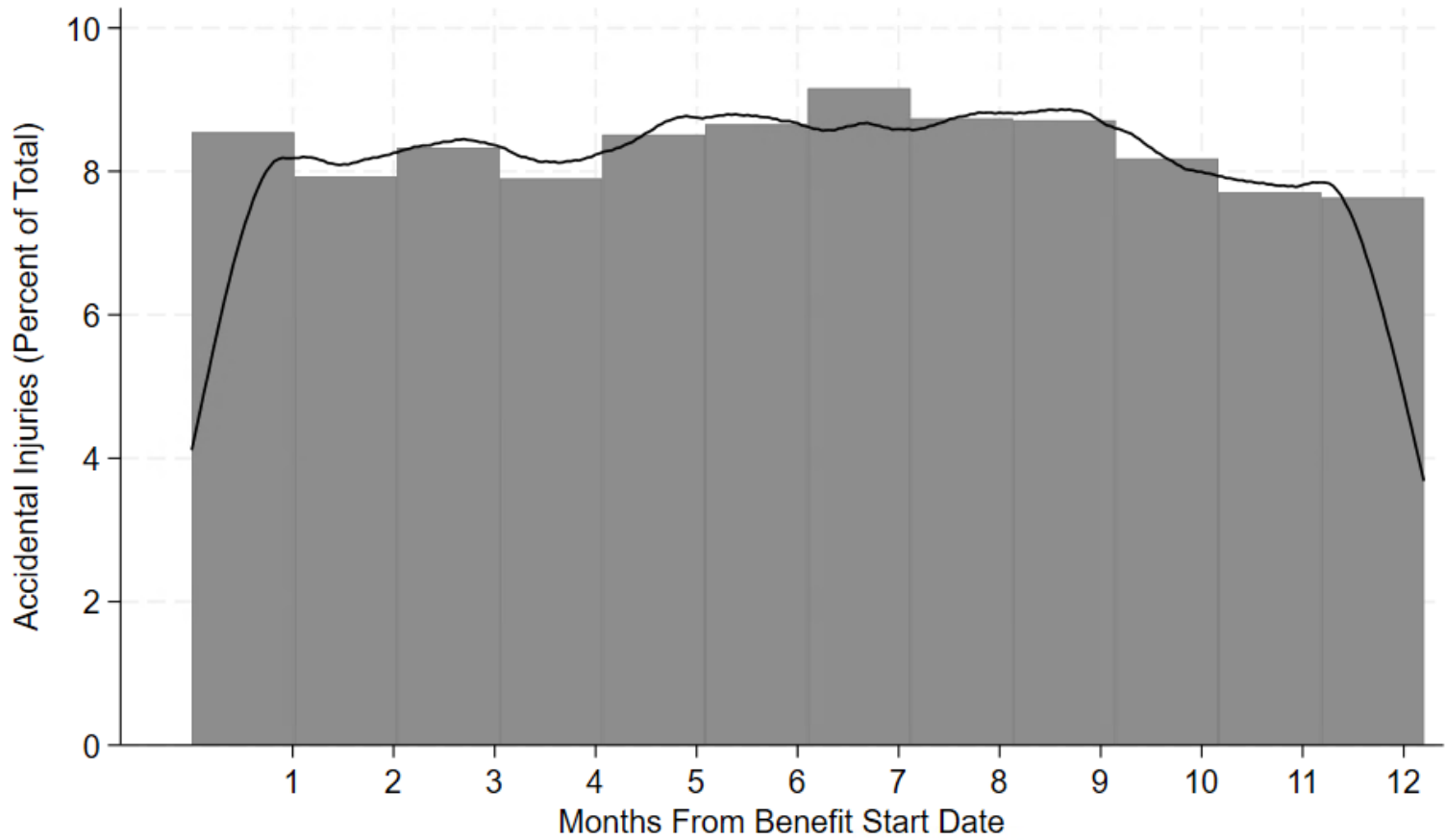

The histogram (gray bars) represents the percent of all accidental injuries that occurred during each month of the benefit period
A kernel density estimate using the Epanechnikov kernel with an optimal half width was added to the histogram to improve visual representation of distribution

**Table A8** Naïve Logistic Model: Marginal Effect of Meeting Family Deductible on Utilization of Healthcare Services

| Type of Care | Obs | M.E. | 95% CI |
|---|---|---|---|
| *Utilization by Setting* | | | |
|     Emergency Department | 269,919 | 3.8 | [3.5 to 4.1] |
|     Inpatient Admissions | 269,919 | 1.9 | [1.7 to 2] |
|     Urgent Care | 269,919 | 3.0 | [2.7 to 3.3] |
|     Outpatient Clinic Visits | 269,919 | 12.5 | [11.9 to 13] |
|     Ambulatory Surgery | 269,919 | 4.1 | [3.9 to 4.4] |
| *Procedures* | | | |
|     Office Visits | 269,919 | 10.1 | [9.5 to 10.6] |
|     Venipuncture | 269,919 | 3.1 | [2.8 to 3.5] |
|     Physical Therapy | 269,919 | 1.8 | [1.6 to 2] |
| *Annual Preventive Care* | | | |
|     Mammography | 24,224 | 1.8 | [0.1 to 3.5] |
|     Wellness Visit | 232,331 | 0.6 | [0.2 to 1.1] |

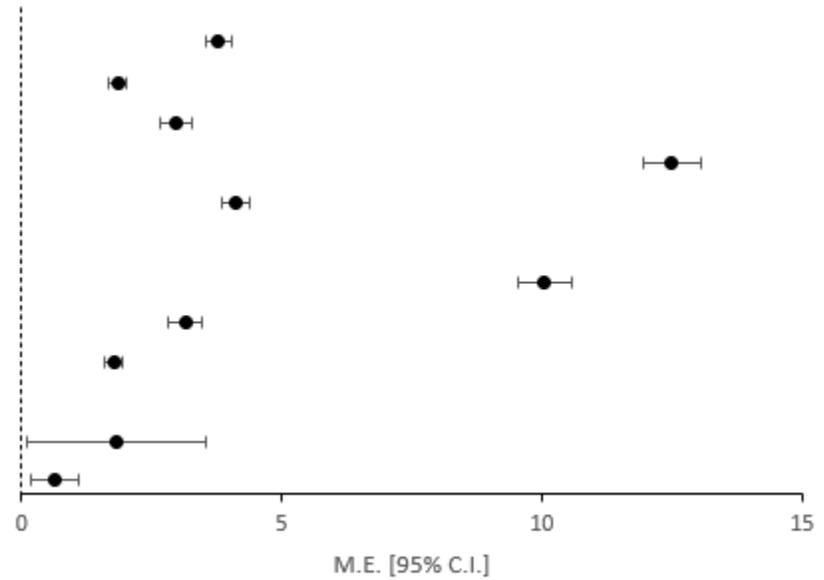

Marginal effects (M.E.) were calculated at mean values for all other predictor variables
95% confidence intervals are presented in brackets to the right of the marginal effect estimates
Total observations (Obs) is presented for each model estimate
All levels of deductible were included in naïve models